\documentclass[preprint]{elsarticle}
\usepackage{amssymb}
\usepackage{graphicx}

\journal{Physics Letters B}

\begin{document}

\begin{frontmatter}
\title{A generalized Heckmann-Schucking cosmological solution in the presence of a negative cosmological constant}
\author{Alexander Y. Kamenshchik\corref{cor1}}
\address{Dipartimento di Fisica and INFN, Via Irnerio 46,40126 Bologna,
Italy\\
L.D. Landau Institute for Theoretical Physics of the Russian
Academy of Sciences, Kosygin str. 2, 119334 Moscow, Russia}\ead
{Alexander.Kamenshchik@bo.infn.it}
\author{Chiara M. F. Mingarelli}
\address{Dipartimento di Astronomia, Via Ranzani 1,40126 Bologna,
Italy}\ead{chiara.mingarelli@studio.unibo.it}

\cortext[cor1]{Corresponding author}
\begin{abstract}
An exact solution of the Einstein equations for a Bianchi -I universe
in the presence of dust, stiff matter and a negative cosmological constant, generalising the well-known Heckmann-Schucking solution is presented.
This solution describes a universe existing during a finite period of cosmic time, where the beginning and the end of its evolution are characterized by
the presence of Kasner type cosmological singularities.
\end{abstract}

\end{frontmatter}
\section{Introduction}
The construction of exact cosmological solutions of the Einstein equations always represents a
rather attractive task. In our opinion  there are now two most important classes
of cosmological solutions. One of them is the class of exact solutions of Friedmann-Robertson-Walker isotropic cosmological
models, which constitutes a basis for comparison of theoretical
predictions with observations \cite{Land-Lif}.
The anisotropic Kasner solution for the empty
Bianchi-I universe \cite{Kasner} is the oldest time dependent cosmological solution of the Einstein equations. Its importance for theoretical
physics is connected with its role in the description of the oscillatory
approach to the cosmological singularity \cite{BKL}, which in turn
appears to be a very promising topic for study in the string and M-theory
context \cite{string}.

The Heckmann - Schucking \cite{Schuck} anisotropic solution for the  Bianchi-I universe in the presence of dust-like matter constitutes a bridge between these two types of the cosmological solutions: in the vicinity of the cosmological
singularity it behaves as a Kasner universe, while at the later stage of
the cosmological evolution it behaves as an isotropic flat Friedmann
Universe. The Heckmann-Schucking solution has the following form:
for the Bianchi-I universe with the metric
\begin{equation}
ds^2 = dt^2 - a^2(t)dx^2 - b^2(t)dy^2 - c^2(t)dz^2
\label{metric}
\end{equation}
filled with dust whose equation of state is
\begin{equation}
p = 0
\label{dust}
\end{equation}
the functions $a(t), b(t)$ and $c(t)$ are given by the formulae
\begin{eqnarray}
&&a(t) = a_0t^{p_1}(t+t_0)^{2/3-p_1},\nonumber \\
&&b(t) = b_0t^{p_2}(t+t_0)^{2/3-p_2},\nonumber \\
&&c(t) = c_0t^{p_3}(t+t_0)^{2/3-p_3},
\label{Schuck}
\end{eqnarray}
where the exponents $p_1, p_2$ and $p_3$ are the well-known Kasner exponents \cite{Land-Lif,BKL}
satisfying relations
\begin{equation}
p_1 + p_2 + p_3 = 1,
\label{Kasner}
\end{equation}
\begin{equation}
p_1^2 + p_2^2 + p_3^2 = 1.
\label{Kasner1}
\end{equation}
Usually the Kasner exponents are arranged in such a way that
\begin{equation}
p_1 \leq p_2 \leq p_3.
\label{order}
\end{equation}

It is easy to see that the solution (\ref{Schuck}) is close to the Kasner solution when $t \ll t_0$. In the limit $t \gg t_0$ all the functions
$a(t), b(t)$ and  $c(t)$ are proportional to $t^{2/3}$, i.e. their behavior
 coincides with that of a flat Friedmann Universe filled with dust.
The generalization of the Heckmann-Schucking solution for the case of Bianchi-I Universe filled with an arbitrary fluid with the equation of state $p = w\rho,\ w = const$, where $p$ is the pressure and $\rho$ is the energy density  was obtained in paper \cite{Jacobs}.

In a recent paper \cite{KhalKam} the Heckmann-Schucking solution was generalized in the case of the Bianchi-I Universe filled with a mixture of three perfect fluids: dust, stiff matter with the equation of state $p = \rho$,  and
a positive cosmological constant.
The consideration of a positive cosmological constant in combination with other types of matter was motivated by the recent discovery of the phenomenon of cosmic acceleration \cite{cosmic}, which  makes the presence of the cosmological
constant or some other type of matter which mimics it necessary
for a realistic cosmological model.  Another argument in favour of a positive cosmological constant  is the fact that
inflationary theories for a very early Universe \cite{inflation} contain
an effective cosmological constant providing a period of a quasi-exponential expansion at the beginning of the cosmological evolution.

However, the consideration of the cosmological models with a negative cosmological constant has some attractive features as well. First of all, the presence of a negative cosmological constant is compatible with  supersymmetry
\cite{supersym}. Secondly, one can formulate various cosmological models, compatible with modern observations, where
the cosmological constant has a negative sign and the present cosmic acceleration is driven by other types of dark energy. In such models, the present accelerated expansion is followed by contraction of the Universe culminating in the Big Crunch cosmic singularity \cite{crunch}. Consideration of the Bianchi-I Universe filled with a negative cosmological constant is therefore interesting because in this case, one has two singular regimes: the Big Bang and the Big Crunch, both characterized by strong anisotropy. In this paper we construct an exact cosmological solution
for a Universe filled with a mixture of three perfect fluids: a negative cosmological constant, dust and stiff matter and study its properties. Section 2 is devoted to the construction of the solution, while the third section contains some conclusions.

\section{Bianchi-I universe in the presence of a negative cosmological constant}
We shall look for solutions of the Einstein equations for the Bianchi - I Universe with the metric (\ref{metric}) filled with three perfect fluids: dust, stiff matter and a negative cosmological constant, whose energy-momentum tensor has the form
\begin{equation}
T_{\mu}^{\nu} = diag(\rho, -p, -p, -p).
\label{EMT}
\end{equation}
We represent the functions $a(t), b(t)$ and $c(t)$ in the following form:
\begin{eqnarray}
&&a(t)  = R(t)\exp(-2\alpha(t)), \nonumber \\
&&b(t) = R(t)\exp(\alpha(t) - \beta(t)), \nonumber \\
&&c(t) = R(t)\exp(\alpha(t) + \beta(t)),
\label{assym}
\end{eqnarray}
where $R(t)$ is the conformal factor while the functions $\alpha(t)$ and $\beta(t)$ characterise the anisotropy of the model. The components of the Ricci tensor have the following form:
\begin{equation}
R_0^0 =-\left(3 \frac{\ddot{R}}{R} + 6 \dot{\alpha}^2 + 2\dot{\beta}^2\right),
\label{Ricci0}
\end{equation}
\begin{equation}
R_1^1 =-\left( \frac{\ddot{R}}{R} + 2\frac{\dot{R}^2}{R^2} -6\frac{\dot{R}}{R}\dot{\alpha}
- 2\ddot{\alpha} \right),
\label{Ricci1}
\end{equation}
\begin{equation}
R_2^2 =-\left( \frac{\ddot{R}}{R} + 2\frac{\dot{R}^2}{R^2} + 3\frac{\dot{R}}{R}(\dot{\alpha} - \dot{\beta})
+ \ddot{\alpha} - \ddot{\beta}\right),
\label{Ricci2}
\end{equation}
\begin{equation}
R_3^3 ==-\left( \frac{\ddot{R}}{R} + 2\frac{\dot{R}^2}{R^2} + 3\frac{\dot{R}}{R}(\dot{\alpha} + \dot{\beta})
+ \ddot{\alpha} + \ddot{\beta}\right).
\label{Ricci3}
\end{equation}
Using the isotropy of the energy-momentum tensor (\ref{EMT}) one has
\begin{equation}
R_1^1 = R_2^2 = R_3^3.
\label{Einstein}
\end{equation}
From this equation it is easy to obtain equations for the functions $\alpha(t)$ and $\beta(t)$:
\begin{equation}
\ddot{\alpha} + 3\frac{\dot{R}}{R}\dot{\alpha} = 0,
\label{alpha}
\end{equation}
\begin{equation}
\ddot{\beta} + 3\frac{\dot{R}}{R}\dot{\beta} = 0.
\label{beta}
\end{equation}
From Eqs. (\ref{alpha}), (\ref{beta}) it follows immediately that
\begin{equation}
\dot{\alpha} = \frac{\alpha_0}{R^3},
\label{alpha1}
\end{equation}
\begin{equation}
\dot{\beta} = \frac{\beta_0}{R^3},
\label{beta1}
\end{equation}
where $\alpha_0$ and $\beta_0$ are some positive constants.

The $00$ component of the Einstein equations now has the form
\begin{equation}
\frac{\dot{R}^2}{R^2} = \dot{\alpha}^2 + \frac{\dot{\beta}^2}{3}
+ \frac{4\pi G}{3}\rho.
\label{Einstein0}
\end{equation}
Choosing a convenient normalization of the constants one can represent
Eq. (\ref{Einstein0}) for a Universe filled with our mixture of three fluids in the following form:
\begin{equation}
\frac{\dot{R}^2}{R^2} = \dot{\alpha}^2 + \frac{\dot{\beta}^2}{3}
- \Lambda + \frac{M}{R^3} + \frac{S}{R^6},
\label{Friedmann}
\end{equation}
where $-\Lambda$ is a negative cosmological term. For convenience we use $\Lambda > 0$,
while the constants $M$ and $S$ characterise the quantity of dust and of stiff matter in the Universe, respectively.

After a substitution into Eq. (\ref{Friedmann}) the expressions for $\dot{\alpha}$ and $\dot{\beta}$ from Eqs. (\ref{alpha1}) and (\ref{beta1}) we come to the following equation for the conformal factor $R(t)$:
\begin{equation}
\frac{\dot{R}^2}{R^2} =
- \Lambda + \frac{M}{R^3} + \frac{S_0}{R^6},
\label{Friedmann1}
\end{equation}
 where
\begin{equation}
S_0 = S + \alpha_0^2 + \frac{\beta_0^2}{3}.
\label{tildeS}
\end{equation}
Notice that the coefficient $S_0$ includes contribution of both the anisotropy
and of the presence of stiff matter. The fact the effective contribution of anisotropy
to the effective Friedmann equation for the conformal factor behaves like that of stiff matter
was first discussed in Ref. \cite{BK}. 
The result of the explicit integration of Eq. (\ref{Friedmann1}) is
\begin{equation}
R^3(t) = \frac{M}{2\Lambda} + \sqrt{\frac{S_0}{\Lambda}}\sin(3\sqrt{\Lambda}t)-\frac{M}{2\Lambda}\cos(3\sqrt{\Lambda}t).
\label{Friedmann2}
\end{equation}
Here the integration constant is chosen in such a way to provide the initial condition
\begin{equation}
R^3(0) = 0.
\label{init}
\end{equation}
It is convenient to also have another form for the solution (\ref{Friedmann2}):
\begin{equation}
R^3(t) = \frac{M}{2\Lambda} + \frac{\sqrt{4S_0\Lambda + M^2}}{2\Lambda}\sin\left[3\sqrt{\Lambda}t
-{\rm arcsin}\left(\frac{M}{\sqrt{4S_0\Lambda+M^2}}\right)\right].
\label{Friedmann3}
\end{equation}
It is seen from Eq. (\ref{Friedmann3}) that the conformal factor $R(t)$, which is equal to zero at the initial moment (Big Bang) expands until some maximum value and then begins contracting, vanishing at some moment $t_{BC}$ (Big Crunch singularity). The value of $t_{BC}$ is
\begin{equation}
t_{BC} = \frac{1}{3\sqrt{\Lambda}}\left[\pi + 2{\rm arcsin}\left(\frac{M}{\sqrt{4S_0\Lambda+M^2}}\right)\right].
\label{BC}
\end{equation}

Now, looking at formula (\ref{BC}) we see that the time of existence of the Universe increases with the growth of the quantity of dust, characterized by the constant $M$ and it diminishes with the growth of the quantity of the effective stiff matter $S_0$. However, changing  both these coefficients one can have variation of the time $t_{BC}$ only up to factor 2. The
dependence of $t_{BC}$ on the negative cosmological constant is much more significant. The time of existence of the Universe decreases with the growth of the absolute value of the cosmological constant and can acquire any value from $0$ to $\infty$.

Substituting the expression (\ref{Friedmann2}) into Eq. (\ref{alpha1}) we can integrate the last equation to obtain an explicit expression for the anisotropy function $\alpha(t)$:
\begin{equation}
\alpha(t) = \frac{\alpha_0}{3\sqrt{S_0}}\ln \left[\frac{\frac{2\sqrt{S_0}}{3\Lambda t_{\alpha}}\sin
\frac{3\sqrt{\Lambda}t}{2}}{\frac{M}{2\Lambda}\sin
\frac{3\sqrt{\Lambda}t}{2} + \sqrt{\frac{S_0}{\Lambda}}\cos
\frac{3\sqrt{\Lambda}t}{2}}\right].
\label{alpha2}
\end{equation}
Here the integration constant is chosen in such a way to provide a Kasner-type behaviour of the function $\alpha(t)$ in the neighbourhood of the Big Bang singularity. Indeed, one can see that
\begin{equation}
\lim_{t \rightarrow 0} \alpha(t) = \frac{\alpha_0}{3\sqrt{S_0}} \ln \frac{t}{t_{\alpha}}.
\label{alpha3}
\end{equation}
The constant $t_{\alpha}$ is introduced into the expression (\ref{alpha2})
 to make the argument of the logarithm dimensionless. The expression for the anistropy function $\beta(t)$ has an analogous structure:
\begin{equation}
\beta(t) = \frac{\beta_0}{3\sqrt{S_0}}\ln \left[\frac{\frac{2\sqrt{S_0}}{3\Lambda t_{\beta}}\sin
\frac{3\sqrt{\Lambda}t}{2}}{\frac{M}{2\Lambda}\sin
\frac{3\sqrt{\Lambda}t}{2} + \sqrt{\frac{S_0}{\Lambda}}\cos
\frac{3\sqrt{\Lambda}t}{2}}\right].
\label{beta2}
\end{equation}

Substituting the expressions (\ref{alpha2}), (\ref{beta2}) and (\ref{Friedmann2}) into Eqs.
(\ref{assym}) we obtain
\begin{eqnarray}
&&a(t)=\left(\frac{\sqrt{S_0}}{3\Lambda t_{\alpha}}\right)^{-\frac{2\alpha_0}{3\sqrt{S_0}}}
\cdot\left(2\sin\frac{3\sqrt{\Lambda}t}{2}\right)^{(\frac13 - \frac{2\alpha_0}{3\sqrt{S_0}})}
\nonumber \\
&&\cdot\left(\frac{M}{2\Lambda}\sin
\frac{3\sqrt{\Lambda}t}{2} + \sqrt{\frac{S_0}{\Lambda}}\cos
\frac{3\sqrt{\Lambda}t}{2}\right)^{(\frac13 + \frac{2\alpha_0}{3\sqrt{S_0}})},
\nonumber \\
&&b(t)=\left(\frac{\sqrt{S_0}}{3\Lambda t_{\alpha}}\right)^{\frac{\alpha_0}{3\sqrt{S_0}}}
\left(\frac{\sqrt{S_0}}{3\Lambda t_{\beta}}\right)^{-\frac{\beta_0}{3\sqrt{S_0}}}
\cdot\left(2\sin\frac{3\sqrt{\Lambda}t}{2}\right)^{(\frac13 + \frac{\alpha_0-\beta_0}{3\sqrt{S_0}})}
\nonumber \\
&&\cdot\left(\frac{M}{2\Lambda}\sin
\frac{3\sqrt{\Lambda}t}{2} + \sqrt{\frac{S_0}{\Lambda}}\cos
\frac{3\sqrt{\Lambda}t}{2}\right)^{(\frac13 - \frac{\alpha_0-\beta_0}{3\sqrt{S_0}})},
\nonumber \\
&&c(t)=\left(\frac{\sqrt{S_0}}{3\Lambda t_{\alpha}}\right)^{\frac{\alpha_0}{3\sqrt{S_0}}}
\left(\frac{\sqrt{S_0}}{3\Lambda t_{\beta}}\right)^{\frac{\beta_0}{3\sqrt{S_0}}}
\cdot\left(2\sin\frac{3\sqrt{\Lambda}t}{2}\right)^{(\frac13 + \frac{\alpha_0+\beta_0}{3\sqrt{S_0}})}
\nonumber \\
&&\cdot\left(\frac{M}{2\Lambda}\sin
\frac{3\sqrt{\Lambda}t}{2} + \sqrt{\frac{S_0}{\Lambda}}\cos
\frac{3\sqrt{\Lambda}t}{2}\right)^{(\frac13 - \frac{\alpha_0+\beta_0}{3\sqrt{S_0}})}.
\label{abc}
\end{eqnarray}	

From Eqs. (\ref{abc}) it follows immediately that at small values of $t$ the scale factors
$a,b$ and $c$ behave as
\begin{eqnarray}
&&a(t) \sim t^{p_1},
\nonumber \\
&&b(t) \sim t^{p_2},
\nonumber \\
&&c(t) \sim t^{p_3},
\label{abc1}
\end{eqnarray}
where
\begin{eqnarray}
&&p_1 = \frac13- \frac{2\alpha_0}{3\sqrt{S_0}},\nonumber \\
&&p_2 = \frac13+ \frac{\alpha_0-\beta_0}{3\sqrt{S_0}},\nonumber \\
&&p_3 = \frac13+ \frac{\alpha_0+\beta_0}{3\sqrt{S_0}}.
\label{abc2}
\end{eqnarray}
It is easy to check that these Kasner indices satisfy the relation (\ref{Kasner}),
while instead of the relation (\ref{Kasner1}) we obtain
\begin{eqnarray}
&&p_1^2 +p_2^2+p_3^2 = 1-q^2,\nonumber \\
&&q^2 = \frac{2S}{3S_0} = \frac{2S}{3\left(S +\alpha_0^2+\frac{\beta_0^2}{3}\right)},
\label{Kasner2}
\end{eqnarray}
where $0 \leq q^2 \leq \frac23$. The relation (\ref{Kasner2}) for the Kasner indices in the presence of stiff matter was obtained in \cite{BK}. In the case $q = 0$ it reduces to the standard formula (\ref{Kasner1}). In this last case it is convenient to introduce the Lifshitz-Khalatnikov  parameter $u$ \cite{LK} such that
\begin{eqnarray}
&&p_1 = -\frac{u}{1+u+u^2},\nonumber \\
&&p_2 = \frac{1+u}{1+u+u^2},\nonumber \\
&&p_3 = \frac{u(1+u)}{1+u+u^2},
\label{LK}
\end{eqnarray}
where $u \geq 1$.

Now, let us consider the behaviour of functions $a,b$ and $c$ in the vicinity of the Big Crunch singularity, i.e. when $t \rightarrow t_{BC}$. It follows from Eqs. (\ref{abc}) that
in this limit
\begin{eqnarray}
&&a(t) \sim  (t_{BC} -t)^{\left(\frac23 - p_1\right)},\nonumber \\
&&b(t) \sim  (t_{BC} -t)^{\left(\frac23 - p_2\right)},\nonumber \\
&&c(t) \sim  (t_{BC} -t)^{\left(\frac23 - p_3\right)}.
\label{abc3}
\end{eqnarray}

It is easy to see that after the transition from a Kasner regime at the beginning of the cosmological evolution ($t \rightarrow 0$) to that at the end of the evolution, the axes,
corresponding to the Kasner exponents $p_1$ and $p_3$ exchange their roles.
At the beginning of the evolution the exponent $p_1$ is the smallest of three exponents
(see Eq. (\ref{order}) ) and at the end of evolution the corresponding exponent $(\frac23 - p_1)$
is the biggest one. Thus, it is convenient to rewrite the relations (\ref{abc}) in the following form:
\begin{eqnarray}
&&a(t) \sim  (t_{BC} -t)^{p'_3},\nonumber \\
&&b(t) \sim  (t_{BC} -t)^{p'_2},\nonumber \\
&&c(t) \sim  (t_{BC} -t)^{p'_1}.
\label{abc4}
\end{eqnarray}
Here we have introduced a new set of primed Kasner exponents, satisfying the relations
(\ref{Kasner}) and (\ref{Kasner2}) and they are ordered in such manner that
$p'_1 \leq p'_2 \leq p'_3$.For the case when the stiff matter is absent ($q = 0$) we can
parametrize the new Kasner exponents by the primed Lifshitz-Khalatnikov parameter $u'$ such that
\begin{equation}
p'_1=-\frac{u'}{1+u'+u'^2},\ p'_2=\frac{1+u'}{1+u'+u'^2},\ p'_1=\frac{u'(1+u')}{1+u'+u'^2}.
\label{LK1}
\end{equation}
It is easy to find the transformation law for the parameter $u$:
\begin{equation}
u' = \frac{u+2}{u-1}.
\label{LK2}
\end{equation}

Here, it is curious to remember that in the process of the oscillatory approach of the Bianchi-IX or Bianchi-VIII Universes to the singularity \cite{BKL,Land-Lif} there are two types of transitions:
the change of a Kasner epoch, i.e. the change of the roles of the scale functions, corresponding to the Kasner exponents $p_1$ and $p_2$  combined with the shift $u' = u-1$ and the change of a Kasner era, when the axes, corresponding to the exponents $p_2$ and $p_3$
swap roles and the parameter $u$ is transformed into $u' = \frac{1}{u}$.

Now, we would like to study the behaviour of the functions $a(t),b(t)$ and $c(t)$ during the whole period $0 < t < t_{BC}$. We shall dwell on the most anisotropic case when the stiff matter is absent. The function $a(t)$ begins its evolution contracting and finishes it also in the contraction phase. Thus, it can have either two extrema (minimum and maximum values) or none.
The function $c(t)$ increases both at the beginning and at the end of the cosmological evolution, hence, it also has either two or no extrema. The function $b(t)$ increases
in the vicinity of the initial singularity and decreases in the vicinity of the final singularity.
It has one maximum value. The extremum condition, which follows from Eqs. (\ref{abc})
can be written down as
\begin{equation}
\sin\left(3\sqrt{\Lambda}t_{i\ ext} + {\rm arcsin}\frac{\sqrt{4\Lambda S_0}}{\sqrt{M^2+4\Lambda S_0}}\right)=(1-3p_i)\frac{\sqrt{4\Lambda S_0}}{\sqrt{M^2+4\Lambda S_0}},\ i=1,2,3.
\label{ext}
\end{equation}

For the function $b(t)$ the exponent $p_2$ satisfies the condition $0 \leq p_2 \leq \frac13$, the absolute value of the expression on the right-hand side of Eq. (\ref{ext}) is less than or equal to one,
and, hence, this equation has always solutions. The only solution such that $t_{2\ ext}$ lies in the interval $(0,t_{BC})$ is the maximum point
\begin{equation}
t_{2\ max} = \frac{1}{3\sqrt{\Lambda}}\left(\pi + {\rm arcsin}\left[(3p_2-1)\frac{\sqrt{4\Lambda S_0}}{\sqrt{M^2+4\Lambda S_0}}\right]-{\rm arcsin}\frac{\sqrt{4\Lambda S_0}}{\sqrt{M^2+4\Lambda S_0}}\right).
\label{ext1}
\end{equation}

For the function $a(t)$ the extremal solutions exist if
\begin{equation}
(1-3p_1)\frac{\sqrt{4\Lambda S_0}}{\sqrt{M^2+4\Lambda S_0}} \leq 1.
\label{extcond}
\end{equation}
Remembering that $-\frac13 \leq p_1 \leq 0$,   one can show that the equation for the extremum has solutions at any acceptable value of $p_1$ provided
\begin{equation}
M^2 \geq 12 \Lambda S_0.
\label{extcond1}
\end{equation}
The last condition (\ref{extcond1}) is quite natural. The appearance of extrema shows that there is a period of isotropization in the evolution of a Universe. In turn, the isotropization is favored by the presence of dust (a large value of $M$) and disfavored by the presence of a negative cosmological constant (the greater is the value of $\Lambda$, the shorter the
time of evolution $t_{BC}$) and by the initial anisotropy of the Universe characterized by the value of the parameter $S_0$.
If the relation (\ref{extcond1}) is broken, then the extrema still exist for the values of
the exponents $p_1$ satisfying the relation (\ref{extcond}). These solutions give the time values corresponding to the minimum and maximum values of the function $a(t)$:
\begin{equation}
t_{1\ min} =
\frac{1}{3\sqrt{\Lambda}}\left( {\rm arcsin}\left[(1-3p_1)\frac{\sqrt{4\Lambda S_0}}{\sqrt{M^2+4\Lambda S_0}}\right]-{\rm arcsin}\frac{\sqrt{4\Lambda S_0}}{\sqrt{M^2+4\Lambda S_0}}\right),
\label{t1max}
\end{equation}
\begin{equation}
t_{1\ max} =
\frac{1}{3\sqrt{\Lambda}}\left(\pi - {\rm arcsin}\left[(1-3p_1)\frac{\sqrt{4\Lambda S_0}}{\sqrt{M^2+4\Lambda S_0}}\right]-{\rm arcsin}\frac{\sqrt{4\Lambda S_0}}{\sqrt{M^2+4\Lambda S_0}}\right).
\label{ext2}
\end{equation}

Similarly the function $c(t)$ has extrema at any acceptable value of the Kasner exponent
$\frac23 \leq p_3 \leq 1$ provided the condition (\ref{extcond1}) is satisfied. In the opposite case the corresponding equation for extrema has solutions for the values of the Kasner exponent $p_3$ such that
\begin{equation}
(3p_3-1)\frac{\sqrt{4\Lambda S_0}}{\sqrt{M^2+4\Lambda S_0}} \leq 1.
\label{extcond2}
\end{equation}
These solutions are
\begin{equation}
t_{3\ max} =
\frac{1}{3\sqrt{\Lambda}}\left(\pi + {\rm arcsin}\left[(3p_3-1)\frac{\sqrt{4\Lambda S_0}}{\sqrt{M^2+4\Lambda S_0}}\right]-{\rm arcsin}\frac{\sqrt{4\Lambda S_0}}{\sqrt{M^2+4\Lambda S_0}}\right),
\label{t3max}
\end{equation}
\begin{equation}
t_{3\ min} =
\frac{1}{3\sqrt{\Lambda}}\left(2\pi - {\rm arcsin}\left[(3p_3-1)\frac{\sqrt{4\Lambda S_0}}{\sqrt{M^2+4\Lambda S_0}}\right]-{\rm arcsin}\frac{\sqrt{4\Lambda S_0}}{\sqrt{M^2+4\Lambda S_0}}\right).
\label{ext3}
\end{equation}

\section{Conclusion}
We have found an exact solution for a Bianchi-I Universe, filled with a mixture of three perfect fluids: dust, stiff matter and a negative cosmological constant. This solution describes the universe existing during a finite period of cosmic time, beginning and ending its evolution in a cosmological singularity. In the vicinity of both of these singularities (Big Bang and Big Crunch) the universe under consideration finds itself in anisotropic Kasner regimes, while during its intermediate evolution it undergoes a kind of isotropization process. The degree of isotropization increases  with the growth of the presence of dust in the universe and decreases with the increasing of the absolute value of the negative cosmological constant. In our opinion, the most interesting feature of this new solution is the fact that it represents the transition from one anisotropic singularity to another. Thus, the generalized Heckmann-Schucking solution studied here can be interpreted as a very simplified model
of a Bianchi-IX Universe, having chaotic oscillatory regimes at the beginning and at the end of its evolution. It would be very interesting to find the relations connecting characteristics of these two regimes, but it could prove to be a rather hard task. To our knowledge the process of evolution away from the singularity with the corresponding transitions from one Kasner regime to another has only been studied numerically (see, e.g. \cite{Henriques}).

\section*{Acknowledgements}
This work was partially supported by the RFBR grant 08-02-00923 and by the grant
LSS-4899.2008.2.

\end{document}